\documentclass{article}
\usepackage{emulateapj,apjfonts}

\def\be{\begin{equation}}
\def\ee{\end{equation}}

\def\dndot{{\Delta \dot N}}
\def\ndot{{\dot N}}
\def\ndotone{{\dot N^{(1)}_{\rm NS}}}
\def\ndottwo{{\dot N^{(2)}_{\rm NS}}}
\def\vgal{{V_{\rm gal}}}
\def\vmax{{V_s}}

\def\porb{{P_{\rm orb}}}
\def\smin{{S_{\rm min}}}
\def\dm{{\rm DM}}
\def\dmax{{D_{\rm max}}}
\def\flp{{f_{L_p}}}

\def\etal{et al.}
\def\A{$\cal A$}
\def\B{$\cal B$}
\def\C{$\cal C$}

\begin{document}
\submitted{Submitted to {\em The Astrophysical Journal}, Nov.\ 18,
1998}
\title{Pulsar Spin Evolution, Kinematics, and the 
Birthrate of Neutron Star Binaries}

\author{Z. Arzoumanian,
	J. M. Cordes,
	I. Wasserman}
\affil{Department of Astronomy, Cornell University, Ithaca, NY
14853-6801}
\affil{arzouman@spacenet.tn.cornell.edu,
cordes@spacenet.tn.cornell.edu,
ira@spacenet.tn.cornell.edu}

\begin{abstract}
From considerations of spin evolution and kinematics in the
Galactic potential, we argue that the pulsars B1913+16,
B1534+12, and B2127+11C may be younger than previously assumed, 
and we find that a lower bound on the formation and merger 
rate of close double neutron star binaries is 
$10^{-6.7} f_b^{-1} f_d^{-1}$ yr$^{-1}$,
where $f_b$ is the beaming fraction and $f_d$ accounts for the
possibility that the known NS-NS binaries are atypical of the
underlying population (e.g., if
most such binaries are born with shorter orbital periods). If we
assume no prior knowledge of the detectable lifetimes of such
systems, the rate could be as large as $\simeq 10^{-5.0} f_b^{-1}
f_d^{-1}$ yr$^{-1}$. From both plausible bounds on $f_b$ and 
$f_d$, and a revision of the independently derived limit proposed
by Bailes (1996), we argue that a firm upper bound is $10^{-4}$ 
yr$^{-1}$.
We also present a unifying empirical overview of the spin-up of
massive binary pulsars based on their distribution in spin
period $P$ and spin-down rate $\dot P$, finding evidence for two
distinct spin-up processes, one dominated by disk accretion, the
other by wind accretion. 
We argue that the positions of binary pulsars in the $P$-$\dot P$
diagram can be understood if
(1)~there exists a Galactic population of pulsars in double 
neutron star systems with combinations of spin and orbital periods 
that have prevented their detection in surveys to date;
(2)~recycled pulsars in wide-orbit binaries are not born near 
the canonical spin-up line in the $P$-$\dot P$ diagram because 
they were predominantly spun up through wind accretion;
and 
(3)~there exists a disfavored evolutionary endpoint for radio
pulsars with spin periods 5--30 ms and $\dot P >
10^{-19}$~s-s$^{-1}$.
\end{abstract}


\section{Introduction}
In anticipation of the new observational window afforded by
interferometric gravitational-wave observatories now under
construction, there has been considerable interest in the rates
of formation and merger of binary systems containing at least
one neutron star. For double neutron-star (NS-NS) systems,
estimates of these rates depend on the assumed ``detectable
lifetimes'' of the radio pulsars in known close NS-NS binaries.

The detectable lifetime of a known NS-NS binary is, presumably,
the sum of the post-accretion age of the observed radio
pulsar and the remaining timespan during which it will continue
to be observable.
For the latter, some past authors have used the
time to coalescence of the system, while others have argued that
luminosity evolution limits the detectable lifetime. For age
estimates, past work has relied exclusively on the
``characteristic'' spin-down timescale $\tau_c = P / 2\dot P$, but
$\tau_c$ is an accurate chronological age only if the
present-day spin period $P$ is much greater than the pulsar's
period $P_0$ at ``birth'' (i.e., at the end of the accretion
phase for recycled pulsars), and for dipolar radiation
braking with a constant magnetic field.  Since $P_0$ may not be 
small compared to $P$, the spindown timescale is generally an
overestimate of the time since spinup; consequently, birth
and merger rates for NS-NS systems based on $\tau_c$ may
be underestimates.

We argue that a neutron star's rotational history since the
cessation of spin-up may be used to better estimate its true age.
For each known system, we check such age estimates
against kinematic constraints derived from  motion in the
galactic gravitational potential (Sections~2 and 3). In
Section~\ref{br.sec}, we use our results to estimate the
detectable lifetimes, and hence birthrates, of the known NS-NS
binaries, revisiting the assumptions underlying earlier work. In
addition, we show that there are kinematic solutions for both
B1534+12 and B1913+16 for which they are young (ages $\sim 10$
Myr); we describe how these ages, if characteristic of binary
neutron stars, might alter estimates of their birth and merger
rates.  Finally, we show in Section~\ref{accret.sec} that the 
phenomena of spiral-in and merger of NS-NS systems are reflected 
in the $P$-$\dot P$ diagram, and argue that details of the spin-up
process for other binary pulsars similarly leave behind
important imprints.

\section{The Ages of NS-NS binaries}

\subsection{The Spin-up Lines}
\label{sulines.sec}
Through accretion of mass and angular momentum from an evolving
companion star, a 
neutron star can be spun up to an equilibrium period approximated
by the Kepler orbital period at the Alfv\'{e}n radius
(Ghosh \& Lamb 1992)\nocite{gl92}, 
$
 P_{\rm eq} = 0.95\,{\rm sec}\;\zeta^{3/2}\,\mu_{30}^{6/7}\,\dot m^{-3/7}\,
 \kappa_{0.4}^{3/7}\,R_6^{-3/7}\,M_{1.4}^{-5/7},
$
where $\mu_{30}$ is the neutron star's magnetic moment expressed 
in units of $10^{30}$ G\,cm$^3$, $\kappa_{0.4}$ is the opacity of
the accreting material in units of 0.4 cm$^2$\,g$^{-1}$
(the Thomson opacity for ionized hydrogen), $R_6$ is the stellar 
radius in units of $10^6$ cm, $M_{1.4}$ is the neutron star's mass in 
units of 1.4 $M_\odot$, and $\dot m$ is the mass accretion rate in units
of the Eddington rate (accretion onto a sphere is assumed). 
The quantity $\zeta \equiv \xi/\omega_s^{2/3}$ is of order unity, where 
$\xi \sim 0.5$ encapsulates uncertainty in the Alfv\'{e}n radius
and $\omega_s \leq 1$ is the fastness parameter. 
In the magnetic dipole model,
$P \dot P = 8\pi^2\eta\mu^2/(3Ic^3)$,
where $I$ is the stellar moment
of inertia and $\eta \lesssim 1$ is a geometrical factor,
which then implies the ``spin-up line''
relationship between $P$ and $\dot P$ for the endpoint of spin-up
by accretion,
\begin{eqnarray}
\label{peqline.eq}
 \dot P & = & \alpha P^{4/3}, \\
 \alpha & = & 1.10\times 10^{-15}\,{\rm s}^{-4/3}
 \;\eta\;\zeta^{-7/2}\;\dot m\;
 \kappa_{0.4}^{-1}\;R_6\;M_{1.4}^{5/3}\;I_{45}^{-1},
\end{eqnarray}
where $P$ is in seconds, and $I = I_{45}10^{45}$ g\,cm$^2$.
\looseness=-1

\medskip
{
\plotone{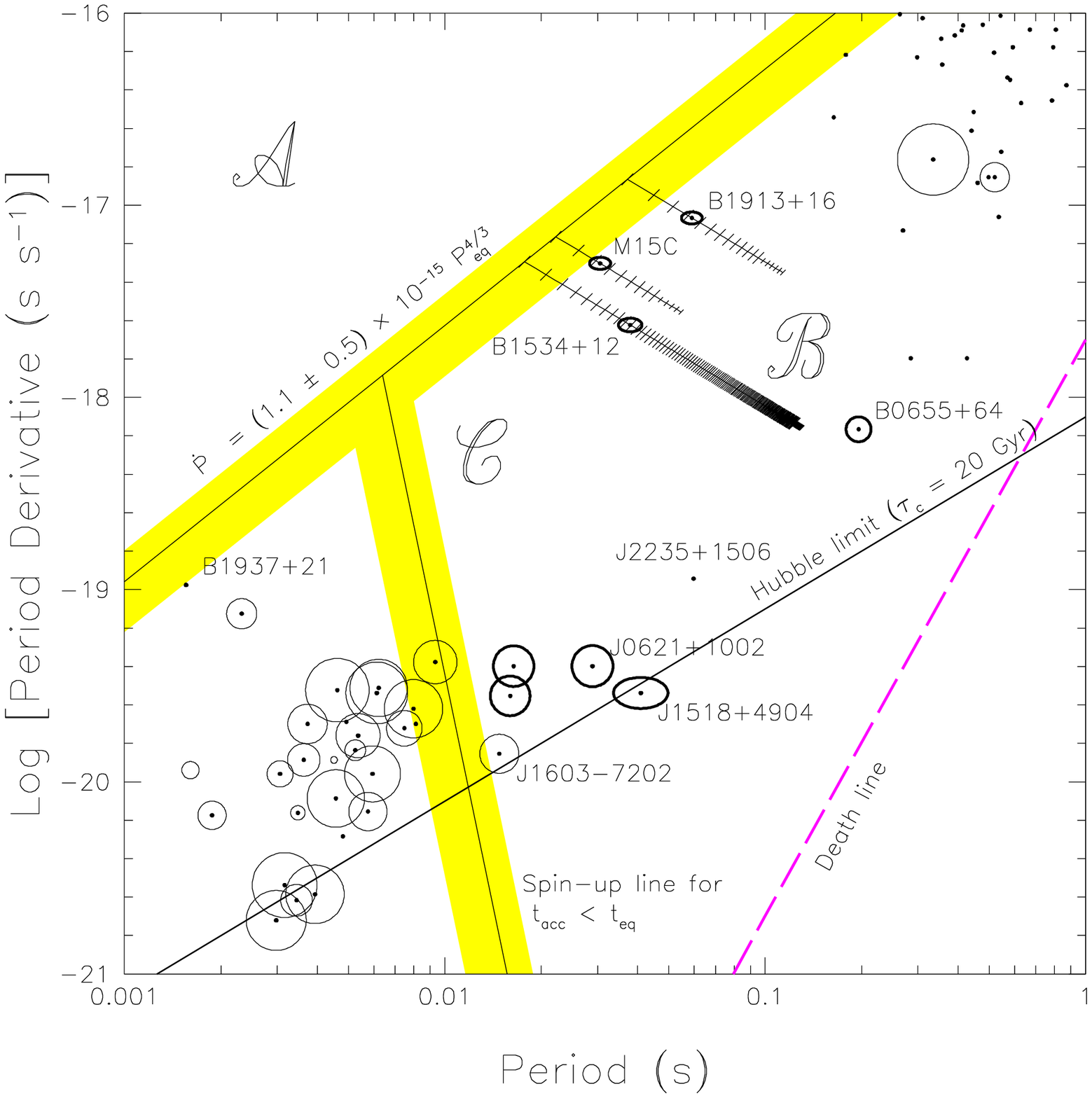} \\
\noindent\parbox{\linewidth}{ \small
{\sc Fig}.~1.---A portion of the $P$-$\dot P$ diagram for pulsars in
the Galactic disk and B2127+11C in the M15 cluster. Pulsars
are shown as dots, with encircled dots denoting binary systems;
the two empty circles represent the eclipsing binaries.
The sizes of the circles scale as $\log P_{\rm orb}$;
heavy circles denote companion masses greater
than 0.45 $M_\odot$, i.e., likely CO rather than He white-dwarf
companions, and ellipses represent double neutron-star
binaries.  Spin-down rates have been corrected for apparent
acceleration due to proper motion
and differential galactic acceleration
(e.g., Camilo, Thorsett \& Kulkarni 1994)\protect\nocite{ctk94}
where such corrections are known.
For PSRs B1534+12, B1913+16, and B2127+11C
(M15C) evolutionary tracks for rotational spin-down assuming
dipolar magnetic braking are shown---tick marks delimit time
intervals of 20 Myr; smaller ticks indicate the portion
of each track for which $P_{\rm orb} < 4$ hours.
The tracks end at the expected coalescence
time for each system, having begun on the equilibrium
spin-up line (eq.~\protect\ref{peqline.eq}), shown as a solid
line within a gray region; the latter indicates the extent to
which the position of this line may vary if assumed
quantities (such as mass accretion rate) take on somewhat
different values.
The spin-up line for $t_{\rm acc} < t_{\rm eq}$,
equation~(\protect\ref{p-8.eq}), assumes $t_{\rm acc} = 10^6$ yr.
The Hubble limit is
the locus of points for which $P/(2\dot P) = 20$ Gyr, and the
death line is the pair-creation limit signifying turn-off of
radio emission.}
}
\medskip

While the scaling $\dot P \propto P^{4/3}$ in equation~(\ref{peqline.eq})
is well-founded for neutron stars with predominantly dipolar magnetic
fields, there are numerous sources of uncertainty in the 
coefficient $\alpha$: the
parameters $\xi$, $\omega_s$, and $\eta$ are not well known, and
sub-Eddington accretion, $\dot m \leq 1$, is likely
(e.g., if inflow onto the neutron star is restricted 
to the polar caps).  Also, the preferred spin-up scenario for NS-NS 
systems, accretion from the helium core of a companion stripped of its
hydrogen envelope,  suggests an
opacity $\kappa_{0.4} = 0.5$. 
We adopt a ``fiducial'' spin-up relation with coefficient
$\alpha = 1.1\times 10^{-15}$~s$^{-4/3}$ and perform calculations
for a range of values.

If the evolutionary timescale of the donor star is short, 
the accretion phase may not last long
enough to spin the neutron star up to equilibrium.
The time to reach equilibrium is 
$t_{\rm eq} \simeq I / (\dot M r^2_A)$;
if mass accretion lasts a time 
$t_{\rm acc} = 10^6t_{\rm acc,6}$ yr $\leq t_{\rm eq}$, 
the spin period at the end of the accretion phase is 
$P = P_{\rm eq} t_{\rm eq} / t_{\rm acc} \geq P_{\rm eq}$,
which implies the relationship
\begin{equation}
\label{p-8.eq}
 \dot P\;P^8_{10} = 3.60\times 10^{-20}\,\eta\;\xi^{-7/2}\;
 \omega_s^{-7}\;\dot m^{-6}\;t_{acc,6}^{-7}\;I_{45}^6\;\kappa_{0.4}^6\;
 R_6^{-6}\;M_{1.4}^{-3},
\end{equation}
where we have used 
$P = 10\,P_{10}$ ms.
Equation~(\ref{p-8.eq}) is extremely sensitive to all of the
parameters, and represents a family of $\dot P \propto P^{-8}$
spin-up lines in the $P$-$\dot P$ diagram; the line
for $\dot m = 1$ and $t_{\rm eq} = 10^6$ yr is plotted in Fig.~1. The
intersection of this alternate spin-up line with the
equilibrium $P^{4/3}$ line defines a minimum post-accretion
period (corresponding to $t_{\rm acc} = t_{\rm eq}$) 
that depends on the total mass accreted, but is insensitive to other
parameters (see also Kulkarni \& Phinney 1994\nocite{kp94}).
If we include the 
effects of accretion-induced field decay following
Shibazaki \etal\ (1989)\nocite{smsn89}, 
we find no qualitative change in the form of the
non-equilibrium spin-up line.

\subsection{Spin-down Ages}

The time to spin down from birth period $P_0$ to present-day period
$P$, with constant braking index $n$, is 
\begin{equation}
\label{age.eq}
  \tau = \tau_c \left[ 1 - (P_0/P)^{n-1} \right],
\end{equation}
where the characteristic age $\tau_c \equiv P/[(n-1)\dot P]$
approaches the true age only in the limit $P_0 \ll P$.
In the $P$-$\dot P$ diagram,
intersection of the spin-down trajectory 
$\dot P = \dot P_0 (P/P_0)^{2-n}$
with an assumed spin-up line provides a
recycled pulsar's initial period which, through equation~(\ref{age.eq})
yields the epoch since the end of spin-up.  We apply this
technique to estimate the ages of the known NS-NS systems;
should accretion in a given system have halted before the
equilibrium spin period was reached, or if the accretion rate
was sub-Eddington, such an estimate becomes a
strong upper bound on the true age.  For the fiducial spin-up
line and $n = 3$, spin-down trajectories for the known NS-NS
systems are shown in Fig.~1, the $P$-$\dot P$
plane for recycled pulsars.  

The validity of post-accretion spin-down ages may be questioned
if there is no single spin-up line as given by
equation~(\ref{peqline.eq}).  Evidence that accretion consistent with
equation~(\ref{peqline.eq}) is responsible for B1913+16-like
systems (see Section~\ref{accret.sec}) is provided by examination 
of the $P$-$\dot P$ diagram:
all three of the known systems lie near the
spin-up line.  The pulsars in these binaries must have
originated above and to the left of their present positions in
this diagram, and in any case
age estimates become increasingly
insensitive to large values of the coefficient $\alpha$
(cf. Fig.~2). 
More importantly, $\alpha$ cannot be made arbitrarily
big, as there are no known recycled pulsars above 
the spin-up line, excepting two 
pulsars in globular clusters (Section~\ref{rega.sec}).
We therefore adopt the value $\alpha = 1.6 \times
10^{-15}$~s$^{-4/3}$ as a reasonable maximum to constrain ages 
at the upper end.

We compute spin-down ages for several values of the braking index,
but for $n \neq 3$ the slope of the spin-up
line differs from $P^{4/3}$ in a way that 
is difficult to quantify without a model for 
the braking torque produced by the magnetic field.
We have applied the model of Melatos (1997)\nocite{mel97}, which 
successfully explains
the measured braking indices $n < 3$ of three young pulsars as a
result of their magnetospheric structure, to objects near the
spin-up line, and find that they should have $n =
3-\epsilon$, where $\epsilon$ is initially very small and
decreases further as objects spin down. This
result supports age estimates derived for 
spin-up lines $\propto P^{4/3}$ and spin-down with $n = 3$. 
If magnetic fields of recycled pulsars evolve
on timescales comparable to $P/\dot P$, however, $n\neq 3$ is possible
(for specific models, see e.g.\ Ruderman, Zhu \& Chen 1998); in 
such models, the spin-down time is altered, but not necessarily
the equilibrium spin-up relation.

\subsection{Kinematic Ages}
\label{kinages.sec}

\looseness-1
A radio pulsar's distance $z$ from the galactic plane together
with its velocity derived from proper motion measurements
can be used to estimate its age through integration of the pulsar's
trajectory in the galactic gravitational potential. 
Sources of uncertainty in addition
to measurement errors in distance and proper motion are the
unknown line-of-sight velocities and non-zero
$z$-heights at ``birth'' (the end of the spin-up phase
for recycled pulsars). We calculate gravitational restoring forces
towards the disk at the present-day galactocentric radii of 
B1534+12 and B1913+16 using the potential of Paczy\'nski 
(1990).\nocite{pac90}

\section{Results for the Known NS-NS binaries}
\subsubsection{PSR B1913+16}

Useful kinematic age constraints can be derived for B1913+16 because
it is at low galactic latitude and its distance and proper motion are
well determined.  Kinematic and spin-down age estimates are 
compared in Fig.~2. Trajectories in the top panel are constrained
to cross the system's current $z$-height moving upward at the
correct present-day velocity $v_z$; they imply that the system
is either leaving the galactic plane for the first time ($\tau
\lesssim 5$ Myr), or has completed at least one-half cycle of
its oscillation in $z$.  In the latter case, i.e., for ages
greater than a few tens of Myr, the system is likely to have
sampled the galactic potential over a large range of
galactocentric radii $r_g$, so that the assumption of a constant
$r_g$ breaks down---this will stretch or compress the
plotted trajectories in a way that depends on the actual path
taken by the system through the Galaxy, which is unknown.  The
unknown line-of-sight velocity $v_r$ also modifies these trajectories, 
but at the system's low galactic latitude, $b = 2^\circ$, 
only a very large $v_r$ would be significant:
to change $v_z$ by 10\%, approximately the difference
between tracks {\em (a)\/} and {\em (b)\/} in Fig.~2, requires 
$v_r \simeq 400$ km-s$^{-1}$.

The characteristic age of B1913+16, assuming a $\dot P$
corrected for galactic acceleration (Damour \& Taylor 1991)\nocite{dt91}, 
is $\tau_c = 109$ Myr, while the combined kinematic and spin-evolution
constraints yield two possible solutions for the age:
a young solution, $\tau \leq 5$ Myr, which is largely
insensitive to uncertainties in the assumed distance,
present-day $v_z$, and braking index, and an old solution,
60--80 Myr,
where the upper bound assumes $n = 3$ and $\alpha \lesssim 1.6\times
10^{-15}$ s$^{-4/3}$. 
It is noteworthy that the old solutions provide an upper bound
on the braking index, $n \leq 3.5$, unless $\alpha$ is very 
large.

\subsubsection{PSR B1534+12}

The characteristic age of PSR B1534+12 is 252 Myr,
assuming the acceleration correction for $\dot P$ given by
Stairs et al. (1998)\nocite{sac+98}. The pulsar's distance from 
the spin-up line
(Fig.~3) implies $\tau \lesssim 210$ Myr, for $n = 3$ and 
$\alpha \lesssim 1.6\times 10^{-15}$ s$^{-4/3}$.  The two estimates
differ little because the condition $P_0 \ll P$ is nearly met if
the pulsar was born on the spin-up line.

Based only on proper motion and distance,
PSR B1534+12 appears to be moving {\em
toward\/} the Galactic plane with transverse velocity $50$
km-s$^{-1}$ from its current position at $z \simeq 800$ pc.
Because the system is at Galactic latitude $b = 48^\circ$, 
a modest radial
velocity could reverse the sign of this apparent motion. Without
more information about the present-day $v_z$ the system's
kinematic history is largely unconstrained.  We have
nevertheless derived the relevant motion in
the Galactic potential under various assumptions.  
In general, the minimum $v_z$ required at birth for the system
to just reach its present $z$-height from a starting point
$z_0 = 0 \pm 200$~pc is $v_z \sim 50$ km-s$^{-1}$.
If we constrain the age to be roughly 200 Myr, allowed
solutions span a range of initial velocities and maximum
$z$-heights depending on the number of oscillation periods that
have  elapsed
since the system's birth, but likely solutions require a birth
$v_z$ between 100 and 200 km-s$^{-1}$ and $z_{\rm max}$ of
2.0--4.5 kpc. 
Notably, the solution in
which the pulsar has completed just less than half of its oscillation
cycle and is returning to the plane for the first time requires
a birth velocity $v_z \sim 275$ km-s$^{-1}$ and $z_{\rm max} = 13$
kpc, which seem unlikely.  If the age is in fact as large as we
believe, the system has probably crossed the plane more than
once.
On the other hand, the kinematic age can be made arbitrarily small
if the radial velocity is assumed to be large and the 
system is leaving the plane for the first time: an age 
$\sim 10$ Myr requires a $v_z$ at birth of 90 km-s$^{-1}$ 
and a present-day $v_r \simeq 140$ km-s$^{-1}$.

\subsubsection{PSR B2127+11C}
The characteristic age of B2127+11C is $\tau_c = 97$ Myr; the
apparent $\dot P$ of this pulsar is likely contaminated by the
system's unknown acceleration in the gravitational potential of
M15 (but Phinney 1992\nocite{phi92b} estimates the contamination
to be at most 2\%).
Although the evolutionary history of B2127+11C is murky,
the pulsar's spin-down track in $P$ and $\dot P$ probably
originated somewhere near the conventional spin-up line.  We
estimate the time since spin-up for this pulsar to be $\sim 60$
Myr (Fig.~3).

\bigskip
{
\plotone{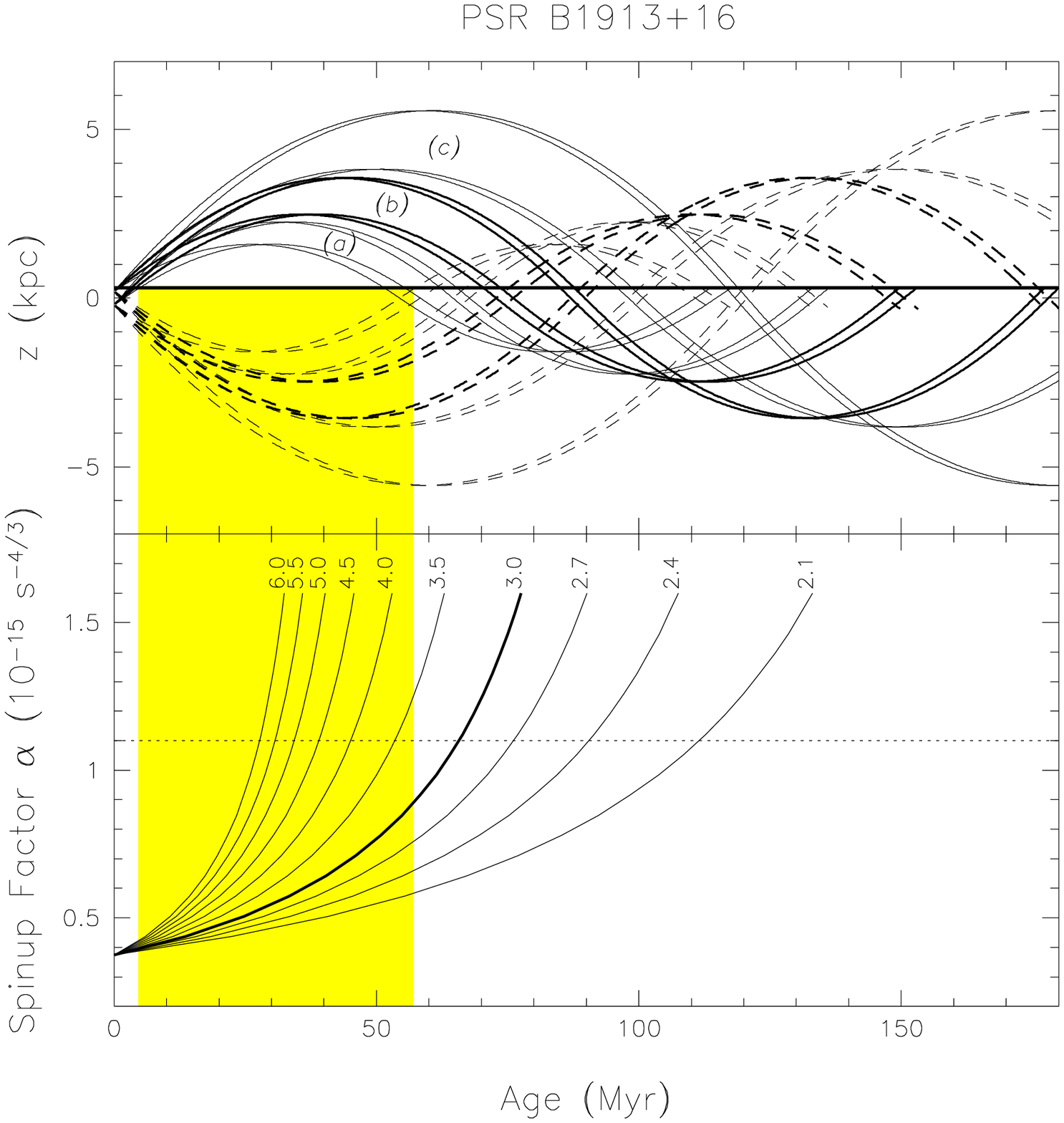} \\
\smallskip
\noindent\parbox{\linewidth}{\small
{\sc Fig.~2.}---Kinematic and spin evolution constraints on the age
of the PSR B1913+16 system. The top panel shows trajectories
perpendicular to the galactic disk allowed by current measurement
uncertainties in distance, $d = 8.3 \pm 1.4$ kpc, and proper motion
(Damour \& Taylor 1991).  The heavy horizontal line represents
the system's present-day height $z = 307 \pm 52$ pc above the
disk.  For each assumed distance, {\em (a)} 6.9 kpc, {\em
(b)} 8.3 kpc (emphasized), and {\em (c)} 9.7 kpc, four
trajectories are plotted, one each for a birth $z$-height of
$\pm 200$ pc and a present-day $v_z$ at either end of the
$1\sigma$ range, $v_z = +(133 \pm 14)(d/8.3\,{\rm kpc})$
km-s$^{-1}$. Dashed curves are mirror-images through $z = 0$ of
the solid curves and show trajectories for negative-going birth
$z$ velocities; for ages between 50 and 120 Myr, these are the
relevant trajectories, as
$v_z > 0$ today.
In the lower panel, the dependence of spin-down age on
the position of the effective spin-up ``line'' is shown for a
variety of braking indices (curves labeled with $n$ at top); the
dotted line indicates the fiducial value of $\alpha$ used in our age
estimates.  The shaded region is kinematically disallowed.}
}

{
\plotone{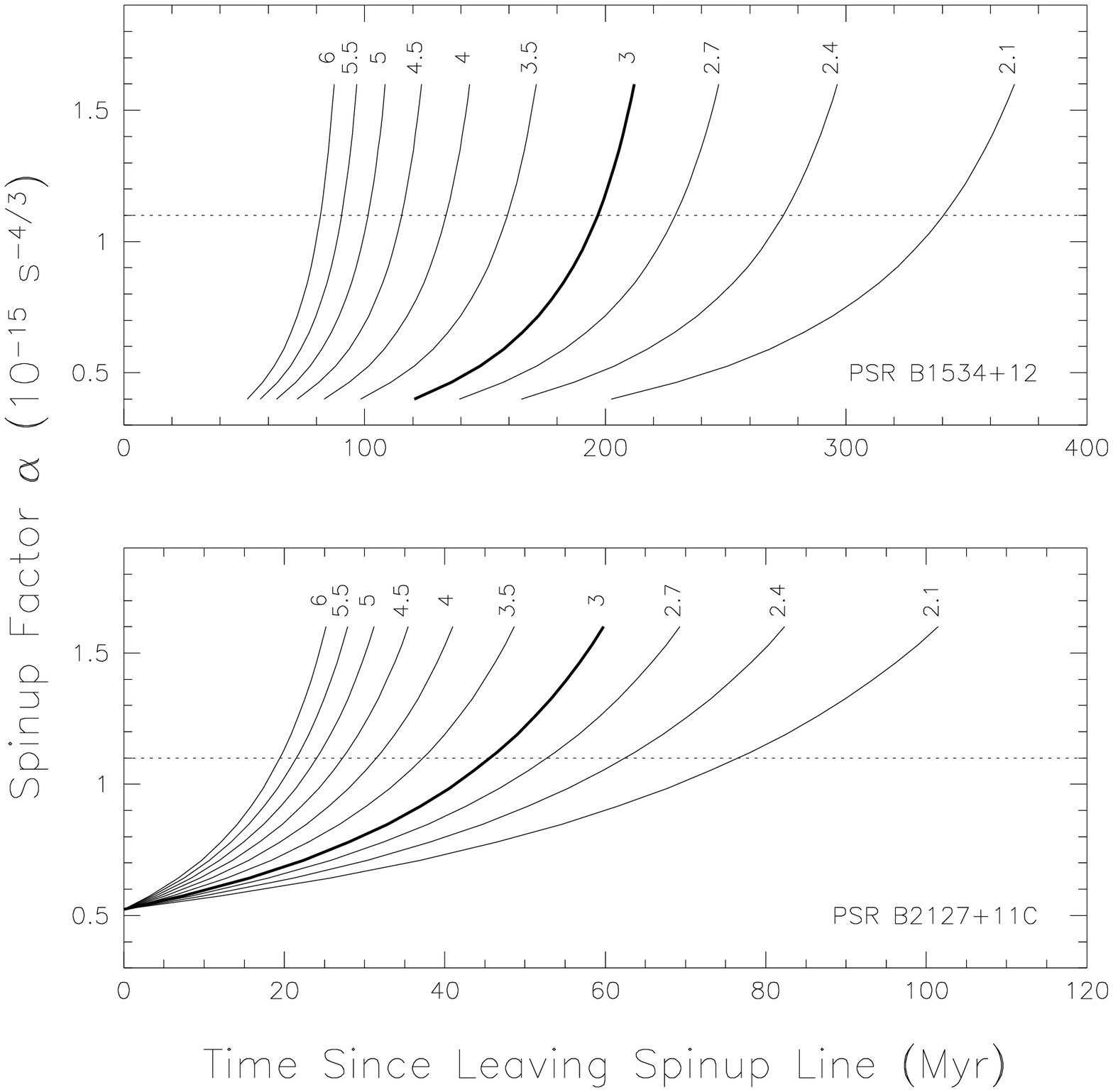} \\
\smallskip
\noindent\parbox{\linewidth}{\small
{\sc Fig.~3.}---Spin-down age constraints for PSRs B1534+12 and B2127+11C.
See caption of Fig.~2 for details.}
}
\medskip

\section{The Formation and Merger Rate of NS-NS Binaries}
\label{br.sec}

\subsection{Maximum Likelihood Birthrate Estimates}
For a given pulsar in a NS-NS binary, the birthrate of similar objects 
may be estimated, following Phinney (1991), using 
$\dndot = \vgal/V_d \tau$, 
where $\vgal$ is the total galactic volume in which NS-NS
binaries are found, $V_d$ is the luminosity- and density-weighted
volume that pulsar surveys
have sampled for objects with identical 
orbital period, and $\tau$ is an appropriate source lifetime.
The total birthrate follows by
summing over objects, $\ndot = \sum_j \dndot_j$. 

A more rigorous treatment, based on a likelihood function, uses all
information in pulsar surveys, including non-detections in much of
the survey volume as well as detected pulsars.  
Although the scarcity of known binary pulsars would frustrate any 
attempt to constrain meaningfully the many important parameters that 
characterize the Galactic population of NS-NS binaries, we can 
nevertheless use the formalism to understand the proper interpretation
of the parameters that enter into any birthrate estimate. We find 
that a maximum likelihood estimate for the binary
neutron star birthrate $\dot N$ is
\begin{equation}
\dot N={N_d\over\langle f\tau\rangle_{exp}},
\label{eq:brest}
\end{equation}
where the number of detected binary pulsars is
$N_d$ and $\langle f\tau\rangle_{exp}$ is the
joint expectation value of the fraction of all
existing binary pulsars detected times source
lifetime, summed over surveys.
The denominator involves an integral of the
number density of binary pulsar systems
(normalized to one) times their average lifetime
convolved with observational selection functions
that limit the spatial volume and range of
lifetimes amenable to detection.
At a more fundamental
level, the necessary integral
involves physical distributions of the positions,
center of mass speeds, orbital elements
and luminosities of binary pulsars at birth,
and also requires physical models for the evolution
of these distributions along with detailed
understanding of the phenomenological limitations
of searches for such systems.

Ideally, the physical
distributions involved in computing
$\langle f\tau\rangle_{exp}$ should be determined
from the available data, but, since very few
NS-NS binaries are known, we, like others before us,
are forced to appeal to prior knowledge or theoretical
expectations about the binary populations instead.
Moreover, we approximate the expectation value
for each survey by
$
\langle f\tau\rangle_{exp}
\approx\langle f\rangle_{exp}\langle\tau\rangle_{exp},
$
where $\langle f\rangle_{exp}$ is the fraction
of all NS-NS binaries we expect to be observable,
and $\langle\tau\rangle_{exp}$ is the detectable lifetime
of such systems (see, e.g., Van den Heuvel \& Lorimer). 
We can also approximate
\begin{equation}
\langle f\rangle_{exp} \simeq
{f_d V_d\over \vgal},
\label{eq:fest}
\end{equation}
where $f_d\leq 1$ is a measure of how representative
the detected binary pulsars are of the
overall Galactic population.
If the NS-NS binaries found so far are typical of the
overall population then $f_d\approx 1$; if not, then
$f_d$ could be small (e.g., a substantial fraction of binary pulsars
could have orbital periods short enough that detection
is unlikely; Johnston \& Kulkarni 1991, Bloom,
Sigurdsson \& Pols 1998). For $f_d=1$, we find a
birthrate estimate from a single detected binary pulsar
similar to Phinney's (1991) $\Delta\dot N$, except
that it is now clear that (i) there is an underlying
assumption that the detected system is typical of
the population and (ii) the appropriate lifetime to
use is the expected time over which the pulsar should
be detectable. Finally, in the absence of detailed knowledge
of the spatial distribution of NS-NS binaries, we shall
approximate (following the language of Cordes \& Chernoff 1997) 
the ``detection volume'' $V_d$ by the ``sampled volume'' $\vmax$,
which is the sensitivity-limited region searched by pulsar surveys.
This approximation is conservative in that $\vmax$ overestimates
$V_d$ (see below).

Apart from $f_d$, a major uncertainty
in any birthrate estimate is the value of
$\langle\tau\rangle_{exp}$. Below, we estimate this
parameter by computing the expected detectable lifetimes
for B1913+16 and B1534+12 separately. How to translate
these values into an estimate of $\langle\tau\rangle_{exp}$
is somewhat subjective, as it depends on one's prior
notions of which, if either, of them is
``more typical.'' This subjectivity pervades previous
estimates of $\dot N$ as well as our own: {\it all\/} are
proportional to nonrigorous estimates of
$\langle\tau\rangle_{exp}^{-1}$. As should be evident
from equation~(\ref{eq:brest}), evaluating $\dot N$
as the sum of birthrate estimates for individual
binary pulsar detections represents a particular
choice for $\langle\tau\rangle_{exp}^{-1}$.

In a manner similar to Phinney (1991), the birth and merger
rates of NS-NS systems have been estimated by Narayan, Piran \&
Shemi (1991) and Curran \& Lorimer (1995) under, effectively,
different assumptions for $\vgal$ and $\vmax$. In each case, the
lifetimes of individual systems were taken to be the sum of the
pulsar's characteristic age and the binary's coalescence time.
Curran \& Lorimer's (1995; hereafter CL95) estimate has twice
been revised. Van den Heuvel \& Lorimer (1996; hereafter VL96)
argued that the effective ``detectable lifetime'' (rather than
the full lifetime culminating in merger) should be used in
birthrate calculations; they estimated radio luminosity decay
timescales for each pulsar and proposed a merger rate nearly
three times that of CL95.  Stairs et al.\  (1998; hereafter
SAC+98) determined that the distance to PSR B1534+12 is larger
than previously believed, implying that $\vmax$ had been
underestimated by CL95. Their correction reduced Van den Heuvel
\& Lorimer's (1996) estimate by a factor of 2.5, bringing it
effectively back to CL95's value. These revisions are summarized
in Table 1.

\begin{table*}
\begin{center}
\normalsize
\caption{Recent (post-1995) birthrate estimates for NS-NS binaries.
For each entry, the 
first item in the pair refers to B1534+12, the second to B1913+16.}
\bigskip
\begin{tabular}{lccccc}
\hline
\hline
&
Scale factor, $\xi$ &
Age, $\tau$ &
Remaining detectable &
Detectable lifetime, &
Individual rate, \\
&
&
(Myr) &
epoch, $\tau_r$ (Myr) &
$T = \tau + \tau_r$ (Myr) &
$\Delta \dot N = \xi/T$ (yr$^{-1}$)\\
\hline
CL95 &
160, 20 & 
250, 109 &
2700, 300 &
2950, 409 &
$5.42\times 10^{-8}$, $4.89\times 10^{-8}$ 
\\
VL96 &
160, 20 &
250, 109 &
510, 220 &
760, 329 &
$2.10\times 10^{-7}$, $6.08\times 10^{-8}$
\\
SAC+98 &
37.8, 20 &
250, 109 &
510, 220 &
760, 329 &
$4.97\times 10^{-8}$, $6.08\times 10^{-8}$
\\
our revision &
37.8, 20 &
200, 65 &
510, 180 &
710, 245 &
$5.32\times 10^{-8}$, $8.16\times 10^{-8}$
\\
\hline
present work$^{a}$ &
$V_{\rm gal}/V_s = 107.1$ &
200, 65 &
510, 180 &
$\bar T = 500$ &
$V_{\rm gal}/V_s\bar T = 2.14\times 10^{-7}$
\\
\hline
\end{tabular}
\end{center}
\normalsize
\noindent $^a$This entry summarizes our estimate $\dot N^{(1)}_{\rm NS}$, 
equation~(\ref{br1.eq}), which is based on different
assumptions for computing galactic scale factors than the first four
entries (see text).
\end{table*}

We first report here our own revision consistent with the
methods and assumptions of CL95.  We then evaluate the
assumptions entering into previous $\vgal$ and $\vmax$
calculations and report a birthrate reflecting recently 
completed radio pulsar surveys.

\subsection{Effects of Revised Lifetimes on Birthrate Estimates}

As argued by Van den Heuvel \& Lorimer (1996), and as emphasized
above, we need the detectable lifetime $T$ of the sources to
estimate their birthrate. VL96 considered radio luminosity decay
and showed that both B1534+12 and B1913+16 will likely become
undetectable before they merge with their companions. 
However, detection is biased against finding recycled pulsars in
short-period binaries; in present
surveys, neither pulsar might have been found if their binary
periods were $\lesssim 0.5$ times their present values. 
The
gravitational radiation lifetimes of B1534+12 and B1913+16 are
$2.7\times 10^9$ yrs and $3.0\times 10^8$ yrs, respectively.
For
B1534+12, the orbital period will halve in about $2.2\times
10^9$ yrs and for B1913+16, in about $1.8\times 10^8$ yrs.  The
latter result implies that the detectable lifetime of B1913+16
is limited by orbital selection rather than by luminosity
decay.  As shown in Table 1, then, our revision to CL95's
estimate consists of adopting (1) the Stairs et al.\ (1998)
distance to B1534+12, (2) our preferred age estimates for
B1913+16, 65 Myr, and B1534+12, 200 Myr, and (3) retaining Van
den Heuvel \& Lorimer's (1996) luminosity decay for B1534+12,
but assuming B1913+16 will remain detectable for 180 Myr as a
result of observational selection due to the rapidly decaying
orbit\footnote{While relativistic precession of the pulsar spin
axis may render B1913+16 undetectable on much shorter timescales
(Kramer 1998, and references therein), for a population of objects 
only the static beaming
properties represented here by the beaming fraction $f_b$ determine
the typical detectable lifetime.}. We thus derive a birthrate, 
uncorrected for beaming or a
luminosity function to account for dim objects, of $1.35\times
10^{-7}$ yr$^{-1}$ by summing the individual rates of the two
objects, a 23\% increase over the latest revision.  We note that
CL95 suggest a factor of 10 increase to account for
low-luminosity objects, and a further factor of three increase
for beaming.

\subsection{Effects of Revised Galactic Volumes on Birthrate Estimates}
Values of $\vgal$ used previously range from $\sim 100$--900
kpc$^3$; the lowest $\vgal$ (CL95) attributes a Population I
radial distribution and a scale height $h_z = 1$ kpc to NS-NS
binaries.  The kinematic ``ages'' of binary pulsars are directly
related to their scale heights. Previously, a range $h_z =
0.5$--1 kpc has been used to estimate binary pulsar birthrates,
but this does not typify the $z$ distribution of old NS which,
for isolated pulsars, includes a large fraction that oscillate
to $z_{\rm max} \sim 10$ kpc.  We argue that the space
velocities of NS-NS binaries are sufficently large (up to $\sim
200$ km-s$^{-1}$) that both the radial and $z$  distributions
should be extended, to roughly a 10 kpc characteristic radius
and $h_z \approx 5$ kpc.  This yields $\vgal \approx 3000$
kpc$^3$. 

Estimates of $\vmax$ are highly sensitive to modeling of selection
effects in surveys that have or could have found the NS-NS binaries. 
The volume sampled (per unit solid angle)
in a survey is (e.g., Cordes \& Chernoff 1997)
\be
\label{vmax1.eq}
\frac{dV_s}{d\Omega_s} = 
\int dL_p\, \flp(L_p)
\int_0^{\dmax} dD\, D^2, 
\ee
where $\dmax = \left(L_p/\smin\right)^{1/2}$ is the maximum distance 
sampled, $\smin$ is the minimum detectable flux density,
and $\flp$ is the luminosity function (with unit area) of the 
``pseudo luminosity'' $L_p = D^2 S_{400}$, where $S_{400}$ is
the 400 MHz flux density tabulated in Taylor, Manchester \& Lyne (1993).  
Equation~(\ref{vmax1.eq}) provides a good estimate of the detection 
volume $V_d$ when the scale height $h_z$ of the NS-NS
population is high enough, that is, when even the brightest pulsars in
the population cannot be detected beyond a vertical distance $h_z$ 
above the location of the solar system in the Galactic disk; for smaller 
$h_z$, equation~(\ref{vmax1.eq}) overestimates $V_d$.
Previous work indicates $\flp\propto L_p^{-2}$ 
over  several orders of magnitude of $L_p$,  for
both isolated pulsars (Lyne et al.\ 1998) and
millisecond pulsars (Cordes \& Chernoff 1997). 
The lower cutoff is taken to be $L_{p_1} \sim 1$ mJy\,kpc$^2$.  For the 
upper cutoff, we use the value for B1913+16,
$L_{p_2} \sim 200$ mJy\,kpc$^2$,    
as its pulse shape suggests that the radio beam points directly
at us, yielding  a large, perhaps maximal luminosity 
for its $P$ and $\dot P$.
This luminosity function does not account for pulsars beamed away 
from us completely, so 
our rates still must be adjusted upward for beaming. 

The minimum detectable flux density,
$\smin$, is generally a strong function of $P$, $\dm$, $\porb$,
and direction, as well as of instrumental parameters.   For
long spin- and orbital-period pulsars,  $\smin$ asymptotes to a 
constant for low $\dm$.
Using the asymptotic $\smin\approx 0.8$ mJy representative
of recent surveys at Arecibo, we obtain
\be
\label{vmax.eq}
\frac{dV_s}{d\Omega_s} < \frac{2}{3}\,S_{\rm min}^{-3/2}\,L_{p_1}\,L_{p_2}^{1/2}
	\approx 10^{-2.4}\,{\rm kpc^3\,deg^{-2}}\;
	S_{\rm min,0.8}^{-3/2}\;
	L_{p_1,1}\;
	L_{p_2,200}^{1/2},
\ee 
where 
$S_{\rm min} = 0.8\,S_{\rm min,0.8}$ mJy, 
$L_{p_1} = 1\,L_{p_1,1}$ mJy\,kpc$^2$, and 
$L_{p_2} = 200\,L_{p_2,200}$ mJy\,kpc$^2$.
Arecibo surveys have covered $\Omega_s \approx 5000$ deg$^2$ (Camilo 
1998)\nocite{cam98}
in recent years; surveys at Parkes and Jodrell have covered more
solid angle but with $\smin \simeq 5$ mJy. 
Together, these imply $\vmax \lesssim 28$ kpc$^3$.
We stress that this value may be a significant overestimate, for two reasons.
First, the asymptotic flux $\smin$ is not appropriate for rapidly rotating,
highly dispersed pulsars, especially at low galactic latitudes. 
Second, the true low-end cutoff of the radio 
luminosity distribution could occur at luminosities much lower than 
1~mJy\,kpc$^2$---Lyne et al.\ (1998)\nocite{lml+98}
show that the luminosity function for
non-recycled pulsars begins to turn over for $L_p \lesssim 10$ mJy\,kpc$^2$,
but that there is no evidence for a turnover at 1~mJy\,kpc$^2$ for spun-up
objects.

Using our estimates of $\vgal$ and $\vmax$, and
expressing $\langle\tau\rangle_{exp}$ in units of the mean of the detectable
lifetimes of B1913+16 and B1534+12, we obtain a rate 
\begin{equation}
\label{br1.eq}
\ndotone=10^{-6.7}\,\,{\rm yr}^{-1}
	\biggl(\frac{\vgal}{3000\,\,{\rm kpc}^3}\biggr)
	\biggl(\frac{500\,\,{\rm Myr}}{\langle\tau\rangle_{exp}}\biggr)
	\biggl(\frac{28\,\,{\rm kpc}^3}{\vmax}\biggr)
	f_b^{-1},
\end{equation}
where $f_b$ is the beaming fraction,
under the assumption that these binaries are typical of the Galactic
population of NS-NS systems.
This result is nearly a factor of 7 smaller than our revision of the CL95
estimate corrected for their suggested factor of 10 contribution from
low-luminosity objects,
$10^{-5.9} f_b^{-1}$ yr$^{-1}$. We attribute this difference to the increased 
galactic volume recently surveyed for rapidly rotating pulsars as
represented here by $\vmax$, and our inclusion of the radio 
luminosity function in estimating $\vmax$ instead of the somewhat
arbitrary multiplying factor of 10 of CL95.

Aside from the beaming factor $f_b$, equation~(\ref{br1.eq}) is a {\em 
lower bound\/} to the actual birthrate for three reasons. First,
the birthrate function for binary neutron stars could be skewed
toward short orbital periods (Bloom, Sigurdsson \& Pols 1998)
where detection becomes unlikely (Johnston \& Kulkarni 1991). If
this is true, then $f_d$ is small, and the birthrate is larger
than $\ndotone$ by a factor of $f_d^{-1}$ (see
eqs.~[\ref{eq:brest}] and [\ref{eq:fest}]).
Second, $\vmax$ overestimates $V_d$: for high-sensitivity surveys such
as those carried out at Arecibo, sky coverage at high latitudes
samples a large volume at $z$-heights beyond the likely scale height of 
NS-NS binaries.
Third, the parameters entering into our estimate of $\vmax$,
equation~(\ref{vmax.eq}), are sufficiently uncertain that the 
sampled volume itself is likely overestimated. We 
have probably overestimated $\langle\tau\rangle_{exp}$ as well.
Taken together, these considerations suggest that $\ndotone$ may
underestimate the true rate by an order of magnitude or more.

Previous estimates of the binary pulsar birthrate, including our
$\ndotone$, presume that their lifetimes are limited by either
spiral-in or spin-down. Ideally, one should dispose with this
prior assumption and determine $\langle\tau\rangle_{exp}$ from 
the data, if possible.
We note that kinematical considerations imply that B1913+16 may be
$\sim 10$ Myr old, and a similarly young age cannot be ruled out for 
B1534+12. If a significant fraction of recycled pulsars
discovered in future surveys prove to be young (age $\sim 10$
Myr), but in binary systems with long gravitational radiation and spindown
lifetimes ($\sim 100$--1000 Myr), then we would be forced to
reduce our estimate of the detectable lifetime $\tau$. (This would 
also imply that something other than gravitational radiation or spindown
regulates the timescale over which binary pulsars are observable.) Such
a short lifetime would require a large birthrate, $\ndottwo \sim
10^{-5.0} f_b^{-1}$ yr$^{-1}$, for such systems, 
which could dominate the rate of
neutron star mergers if they are common. 
As long as there is not a statistically
significant population of anomalously young binary pulsars, this large
birthrate is not required. Given that only a few pulsars are known to
exist in tight binaries, however, we cannot rule out the existence of
short-lived binary pulsars that are born frequently.

Both of our birthrate estimates $\ndotone$ and $\ndottwo$
can easily take on values larger than $10^{-5}$ yr$^{-1}$ when
provision is made for beaming and the large uncertainties in 
$\vgal$, $\vmax$, and $V_d$. This possibility prompts us to re-examine
the upper bound of $10^{-4.9}$ yr$^{-1}$ proposed by Bailes (1996)
based on the absence of any known ``normal'' (i.e., non-recycled, 
long-period, high magnetic-field) pulsars in a close
NS-NS binary. 
We begin by defining the following quantities:
\begin{itemize}
\item $\dot N$, the birthrate of NS-NS binaries that will merge in a 
	    Hubble time,
\item $T_{nb}$, the detectable lifetime of the non-recycled neutron star 
	    in such a binary,
\item $f_{vb}$, the fraction surveyed to date of the galactic volume 
	    occupied by such binaries ($\equiv V_d/\vgal$),
\item $f_{\rm orb}$, the fraction of such binaries that are detectable 
	    despite orbital smearing of the {\em non-recycled\/} 
	    pulsar's pulsations,

\item $\dot N_n$, the birthrate of (normal) isolated pulsars,
\item $T_n$, the detectable lifetime of a normal pulsar,
\item $f_{vn}$, the fraction surveyed to date of the galactic volume 
	    occcupied by normal pulsars,
\item $N_n$, the number of non-recycled pulsars known to date.
\end{itemize}
As no normal pulsars have been found in NS-NS systems 
expected to merge in a Hubble time,\footnote{The NS-NS system B2303+46,
which is known through the non-recycled neutron star, has $P_{\rm orb} = 
12.3$ days and will not coalesce in less than a Hubble time.}
we have
\be
  \dot N \times T_{nb} \times f_{vb} \times f_{\rm orb} \times f_b < 1.
\ee
Similarly, for the normal, isolated pulsar population, 
\be
  \dot N_n \times T_n \times f_{vn} \times f_b = N_n.
\ee
Taking the ratio, and assuming $T_{nb} = T_n$, we get
\be
\label{bailes.eq}
  \dot N < \frac{1}{N_n} \biggl(\frac{f_{vn}}{f_{\rm orb}f_{vb}}\biggr) 
	\dot N_n.
\ee
The quantity in parentheses, which was not considered by Bailes (1996) 
explicitly, can be less than or greater
than one, but is probably close to unity. We believe the
scale height of binaries is somewhat smaller than that of the isolated
normal pulsars, so $f_{vb} \gtrsim f_{vn}$ is likely (i.e., a larger fraction
of the galactic volume of binaries has been searched). Because of orbital
acceleration, $f_{\rm orb}$ is 
less than one, but represents a less severe selection effect for 
slow pulsars than for recycled objects.

Bailes (1996) assumed $N_n = 650$, and adopted the largest allowed
birthrate, $\B_n = 1/125$ yr$^{-1}$, from the analysis by Lorimer et 
al.\ (1993)\nocite{lbdh93}
of the population of normal, isolated pulsars. A
similar recent effort by Lyne et al.~(1998) based on a larger
sample of objects finds a maximum birthrate of one every 60 yrs.
Furthermore, the number of cataloged normal pulsars with
measured spin-down rates, i.e., for which non-membership in a
NS-NS binary has been firmly established, is 501. 
Assuming unity for the factor in parentheses in equation~(\ref{bailes.eq}),
an upper bound on the birthrate of close NS-NS binaries is then
$\dot N < 10^{-4.5}$ yr$^{-1}$,
independent of beaming.
This rate should be corrected upward by a statistical ``Poisson''
factor and by the factor $f_{\rm orb}$; we therefore believe that
a robust upper limit is 
\be
	\dot N < 10^{-4}\,\,{\rm yr}^{-1}.
\ee
We point out, however, that 
the rate $\ndottwo$ need not be subject to this bound, as the assumption
$T_{nb} = T_n$ might not hold if the detectable lifetimes of NS-NS
binaries were somehow not governed by the spin-down or gravitational 
radiation timescales.

Finally, we refer, for completeness, to the recent work of Rosswog et al.\ 
(1998) who find that a NS-NS merger rate of $10^{-5.5}$--$10^{-4.5}$ 
yr$^{-1}$ could acount for the apparent deficit of $r$-process nuclei.

\section{Accretion Mechanisms and the 
Distribution of Recycled Pulsars in $P$ and $\dot P$}
\label{accret.sec}
Surveys over the last 16 years have been sensitive to objects
with $P \gtrsim 1$ ms to varying degrees.  Certain portions of
the $P$-$\dot P$ diagram have become populated as a result,
while others have remained empty. We argue that three of these
unpopulated regions, labelled \A, \B, and \C\ in Fig.~1, provide
insights into the evolution of binary systems containing neutron
stars. In particular, there is a dearth of pulsars in region \B\ 
where we would expect to find inspiralling NS-NS binaries. We assess
the significance of this and other deficits below.

\subsection{Region \A}
\label{rega.sec}

There are no known selection effects to prevent detection of
pulsars in region \A: objects with spin-down rates higher than
the low-field millisecond pulsars would likely be brighter and so
more easily detectable. Still, no recycled pulsars in the
galactic disk are known above the spin-up line.  We interpret this
as a qualitative validation of the Ghosh-Lamb spin-up line.
We discount pulsars in globular clusters, here and in the discussions 
to follow, because of their uncertain spin-down rates and
complicated evolutionary histories.

\subsection{Region \B}
\label{regb.sec}
Region \B\ is below and to the right of the known NS-NS systems, for
$50\,{\rm ms} \lesssim P \lesssim 200\,{\rm ms}$ (Fig.~1).
A deficit of known pulsars here can occur for three
reasons, all of which likely play some role:

\begin{enumerate}
\item NS-NS binaries born near the spin-up line will 
coalesce before reaching region \B. In Fig.~1, the spin-down 
tracks plotted for the three NS-NS systems 
end at lower right at the expected coalescence time
for each system. Only B1534+12 penetrates 
into region \B\ before merging.

\item For systems like B1534+12, inspiral results in short 
orbital periods, and high orbital velocities, within region 
\B, and such pulsars have been selected against.

\item As the pulsars in these systems age, their radio
luminosities probably decay, making them more difficult to
detect (Van den Heuvel \& Lorimer 1996).  
\end{enumerate}

Evidently, any systems in \B\ are selected against in
pulsar surveys, or few live long enough to reach or spend
sufficient time in the region. Apparently, no new systems like B1913+16
are being formed far from the spin-up line.
The other known pulsars in the vicinity of region \B\ are
consistent with this hypothesis. The 0.8 $M_\odot$ white-dwarf
companion to PSR B0655+64, the ``failed NS-NS'' system, has cooling age
$\sim 4$ Gyr (Hansen \& Phinney 1998)\nocite{hp98b}, which suggests the pulsar
emerged from the accretion phase near the spin-up line with
$P_0 \sim 20$ ms. Also, J2235+1506 and two objects with $P > 200$ ms
are solitary and so could not have escaped detection through
merger or orbital acceleration. The suggestion of 
Camilo, Nice \& Taylor (1993)\nocite{cnt93}
that J2235+1506 is the result of a disrupted binary is
consistent with this view.

The deficit of pulsars in region \B\ supports our use
of the spin-up line as a zero-age starting point, and
is consistent with the existence of
a population of NS-NS systems with periods shorter than 8
hours which has remained undetected because of sensitivity
limits and observational selection against binary pulsars with
short orbital periods and massive companions.

\vspace{-5ex}
\subsection{Region \C}
\label{regc.sec}
Pulsars should already be known in region \C, periods 
5--30 ms and $\dot P > 10^{-19}$ s-s$^{-1}$,
where they would have spin-down luminosities
1.5--2 orders of magnitude higher than, say, J0621+1002.  
We assert that the evolutionary path for producing radio pulsars
in region \C\ is disallowed or rarely followed.

A clue to the absence of pulsars in \C\ may lie in the 
apparent bimodality in the spin parameters of
the massive binaries (Fig.~1). The ``B1913'' family
(B1913+16, B1534+12, B2127+11C, and B0655+64) consists of
pulsars in short-period orbits with NS (or white-dwarf)
companions; the pulsars had $P_0 \gtrsim 20$ ms at the
end of the spin-up phase. The ``J0621'' family (J0621+1002,
J1518+4904, J2145$-$0750, and J1022+1001) consists of pulsars in
$\sim 8$-day orbits with massive white-dwarf (or NS) companions.
The division between the two families is strikingly clean---the
systems with discrepant companion types 
(B0655+64 
and J1518+4904) 
nonetheless have orbital periods consistent with their 
respective family members.\footnote{We predict that should the
NS companion to B2303+46, a 1 s pulsar in a 12 day eccentric
orbit, ever be detected as a pulsar, it will lie among the
J0621 family, alongside the similar NS-NS system J1518+4904.}

The shorter average spin period of the J0621 family implies that 
spin-up in the wide-orbit systems must have been at least as
effective as in the close NS-NS binaries,
but produced final magnetic fields smaller by more than
an order of magnitude.  These facts suggest that different
accretion and spin-up processes were responsible: 
we believe that the B1913 family was spun up through disk
accretion and the J0621 family predominantly during a 
wind-driven accretion phase
(e.g., Van den Heuvel 1994\nocite{vdh94} presents a spin-up scenario
for J2145$-$0750 involving wind accretion). It is generally believed
that disk accretion drives monotonic spin-up (until the spin period
$\simeq P_{\rm eq}$) once the donor star overflows its Roche lobe, 
whereas wind accretion produces a non-monotonic random walk in the 
angular momentum imparted to the NS. 
To achieve the same final spin period, then, much more mass must be
transferred, at lower accretion rates, in wind accretion, resulting in smaller
magnetic fields for the wide-orbit binaries; 
this is consistent with viable models of field attenuation during
accretion (Konar \& Bhattacharya 1998)\nocite{kb98}.
(Wijers 1998 refutes the hypothesis that field loss 
scales with accreted mass, but
without considering non-monotonic spinup.) 
Although recent observations show that disk accretion is also not 
entirely monotonic (Bildsten \etal\ 1997; Nelson \etal\
1997\nocite{bcc+97,nbc+97}), and certain numerical simulations of 
wind accretion question the existence of angular momentum 
``flip-flops'' (e.g., Font \& Ibanez 1998)\nocite{fi98},
frequent or deep torque reversals during spin-up in wide
binaries can account for the two distinct families of massive 
binary pulsars without the need for abandoning field decay 
that scales with accreted mass.

Given the proximity of region \C\ to the spin-up line and the implied 
high magnetic fields relative to J0621+1002, our arguments
suggest that the missing NS undergo accretion in close systems.
We speculate, then, that the initial conditions
(orbital separation and companion type) needed to produce
pulsars in region \C\ lead instead to accretion
ending with collapse of the NS to a black hole or merger with its
companion in a common-envelope phase. In detail,
the outcome would depend on the maximum mass of neutron stars and
the efficiency of the accretion (see, e.g., Brown 1995 and Fryer, 
Benz, \& Herant 1996).

Detecting pulsars in region \C\  should also be made difficult
by the same pulse-smearing effects that would prevent detection
of pulsars like B1913+16 in binary orbits with
periods smaller than about four hours (Johnston \& Kulkarni
1991; Section 4 above). The minimum detectable spin period for a
given orbital period is $P_{\rm min}\propto P_{\rm orb}^{-4/3}$,
so it is unlikely that pulsars spinning faster than $P\sim 20$
ms could be detected in binaries with $P_{\rm orb}\approx 8$
hrs.  Conversely, pulsars with $P\sim 20$ ms could only be found
in binaries with orbital periods longer than about 8 hours.
This means that recycled pulsars in binaries like those that
contain B1534+12, B1913+16, and B2127+11C could remain
undiscovered throughout region \C. Moreover, the absence of
pulsars in somewhat longer period binaries (e.g. $P_{\rm
orb}\sim 0.5$--1 day) in region \C\ suggests that there is a
fairly sharp upper cutoff to the distribution of periods of
close binary pulsar systems, the extended B1913 family. This
supports the idea that the much longer period binaries in the
J0621 family have a completely different physical origin.

\subsection{Discussion}

The explanation commonly put forth for millisecond pulsars with
$\tau_c$ exceeding a Hubble time is that these objects were born
with spin periods close to their present values:  spin-up
endpoints are assumed to adhere to equation~(\ref{peqline.eq}), but
with very low accretion rates, $\dot m \sim 10^{-2}$ (e.g.,
Hansen \& Phinney 1998)\nocite{hp98b}. We suggest two additional ways 
in which
spin-up endpoints far from the canonical spin-up line may be produced:
(1) for short-period binaries ($P_{\rm orb} \lesssim 2$ d), 
pulsars emerge on one of the $P^{-8}$ spin-up lines if
accretion ceases before the equilibrium period is reached;
and
(2) for long-period binaries, wind accretion results in low-field
radio pulsars with as yet undetermined initial periods. 
The ages of such systems are therefore uncertain.
It would seem that near-Eddington disk accretion given
sufficient time to reach equilibrium rotation and ending in a
pulsar is rare, as is indicated by the paucity of radio pulsars
near the spin-up line for $P < 30$~ms. In addition, if prolonged
mass transfer with minimal spin-up during the wind accretion phase 
is an important part of the evolution of wide-orbit millisecond
pulsars, as we suggest, the masses of these objects may be 
considerably higher than previously believed. Future high-precision
mass measurements for such systems may therefore provide 
important constraints on the maximum allowed neutron star mass.

\section{Conclusion}
\label{concl.sec}
By estimating the detectable lifetimes of PSRs B1534+12 and
B1913+16, we have revised previous estimates of the galactic
birthrate of close double neutron star binaries. In steady 
state, this would also be the coalescence rate for these
relatively short-lived binaries.  We find, adopting the 
galactic ``scale factors'' of Curran \& Lorimer (1995), that 
the birthrate of such systems, uncorrected for a radio pulsar
luminosity function and beaming, is $10^{-6.9}$ yr$^{-1}$.  Our
own estimate of the birthrate, independent of Curran \&
Lorimer's (1995) methodology, is based on a maximum likelihood
analysis that clarifies the assumptions entering into all
birthrate estimates, including our own. We find, given available
knowledge of the ages, systemic velocities, and orbital
properties of the two known close NS-NS systems in the galactic
disk, along with an estimate of the galactic volume surveyed to
date, that the birthrate (and hence merger rate, for a
steady-state distribution) is $\gtrsim 10^{-6.7} f_b^{-1}
f_d^{-1}$ yr$^{-1}$, where $f_b$ is the beaming factor, and $f_d
\leq 1$ represents the degree to which the two known objects are
typical of the entire population of NS-NS binaries. Important
parameters entering into this estimate, e.g., the lower cutoff
of the luminosity function for recycled pulsars, are
sufficiently uncertain, however, that the true rate is probably 
higher by an order of magnitude or more; we therefore consider
our result to be a realistic lower bound on the true rate. 
Furthermore, we find
that kinematic and spin-down age constraints for PSRs B1534+12
and B1913+16 allow both systems to be unexpectedly young---if we
dispose of prior prejudices about the detectable lifetimes of
these systems, their birthrates could be substantially
higher, of order $10^{-5} f_b^{-1} f_d^{-1}$ yr$^{-1}$. 
We have revised the upper bound on the NS-NS binary birthrate
proposed by Bailes (1996) based on the absence of any known
non-recycled pulsar in a NS-NS binary, finding that the
potentially large birthrates allowed by our analysis are
consistent with a revised upper bound at $10^{-4}$ yr$^{-1}$.
We therefore bracket the merger rate of NS-NS binaries to the
range $10^{-6.7} f_b^{-1} f_d^{-1}$ yr$^{-1} \lesssim \dot N
\lesssim 10^{-4}$ yr$^{-1}$.

Our considerations of $V_{\rm gal}$ for NS-NS binaries suggest that
searches for such binaries out of the Galactic plane will be more
productive than those in the plane; independent of how many are
found, high-latitude searches will allow determination of the 
thickness of the population in $z$.

This work was supported by the NSF and NASA grants AST95-28394 and 
NAG5-3097.

\end{document}